\documentclass[]{iopart}     % end version size

\usepackage{iopams}
\usepackage{graphicx}  

\begin{document}

\title[Transverse momentum spectra and elliptic flow at LHC]
{Transverse momentum spectra and elliptic flow: Hydrodynamics with 
QCD-based equations of state}

\author{M Bluhm$^1$, B K\"ampfer$^{1,2}$ and U Heinz$^3$}

\address{$^1$ Forschungszentrum Dresden-Rossendorf, PF 510119, 01314 Dresden, Germany}
\address{$^2$ Institut f\"ur Theoretische Physik, TU Dresden, 01062 Dresden, Germany}
\address{$^3$ Department of Physics, The Ohio State University, Columbus, OH 43210, USA}

%\ead{m.bluhm@fzd.de}

\begin{abstract}
  We present a family of equations of state 
  within a quasiparticle model adjusted to lattice QCD 
  and study the impact on azimuthal flow anisotropies and transverse momentum spectra 
  within hydrodynamic simulations for heavy-ion collisions at energies relevant for LHC. 
\end{abstract}

%insert if necessary:
%\pacs{12.38.Mh;25.75-q;25.75.Ld}
%\submitto{\JPG}

%\maketitle  %if used a newpage command will be applied and forces a pageturn

\section{Introduction}

The equation of state (EoS) represents the heart of hydrodynamic simulations for ultra-relativistic 
heavy-ion collisions. Here, we present a realistic EoS for QCD matter delivered by our 
quasiparticle model (QPM) faithfully reproducing lattice QCD results. 
The approach is based 
on~\cite{Peshier:1994zf,Peshier:1995ty,Peshier:1999ww,Peshier:2002ww,Bluhm:2006yh} 
adjusted to the pressure $p$ and energy density $e$ of $N_f=2+1$ quark 
flavors~\cite{Karsch:2003zq,Karsch:2003vd}. As the QPM EoS does not 
automatically fit to the hadron resonance gas EoS in the confinement region, we construct a family of EoS's by an interpolation between the hadron resonance gas at 
$e_1=$ 0.45 GeV/fm$^3$ and the QPM at flexible $e_m$ (cf.~\cite{Bluhm:2007nu} for details). 
In this way, the influence of details in the transition region on hydrodynamic flow can 
be studied, since for $e<e_1$ and $e>e_m$ the EoS is uniquely given by the resonance gas 
and the QCD-based QPM, respectively. In Figure 1, 
we exhibit the EoS family in the form $p=p(e)$ and the corresponding speed of sound $v_s^2=\partial p / \partial e$. For LHC, baryon density effects are negligible. 
\begin{figure}[h]
  \hspace{0.3cm}
  \includegraphics[scale=0.51]{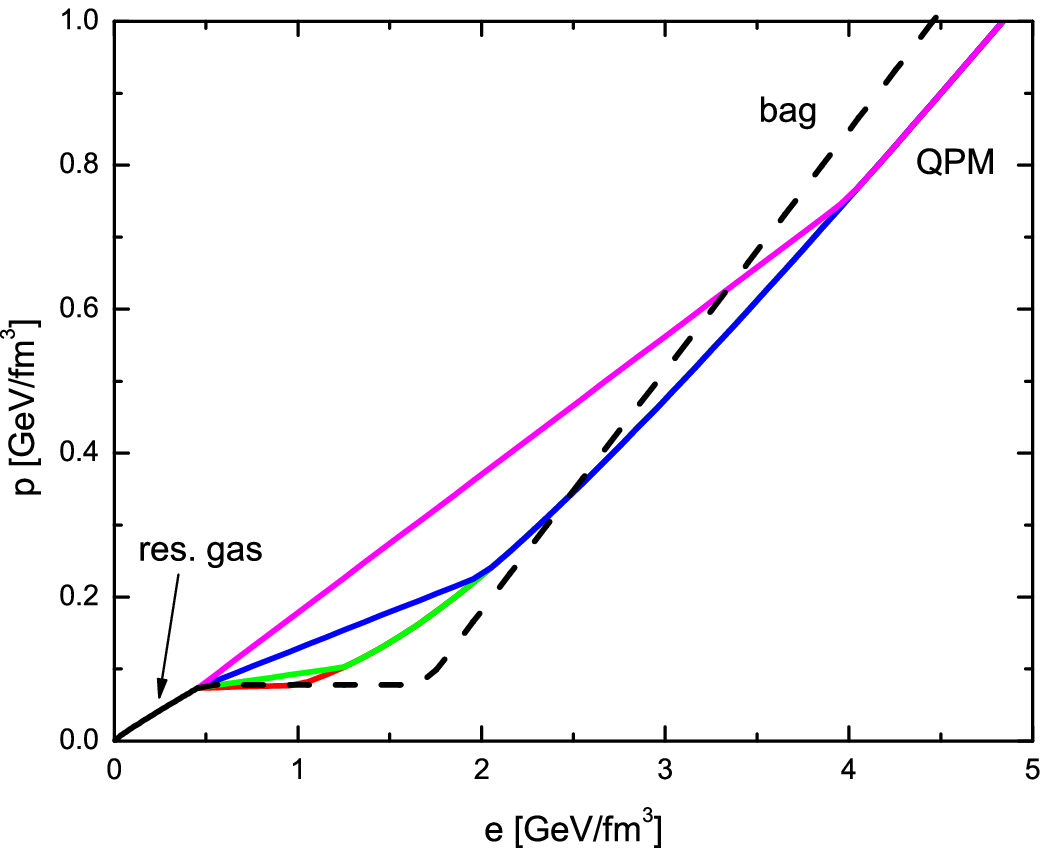}
  \hspace{0.7cm}
  \includegraphics[scale=0.515]{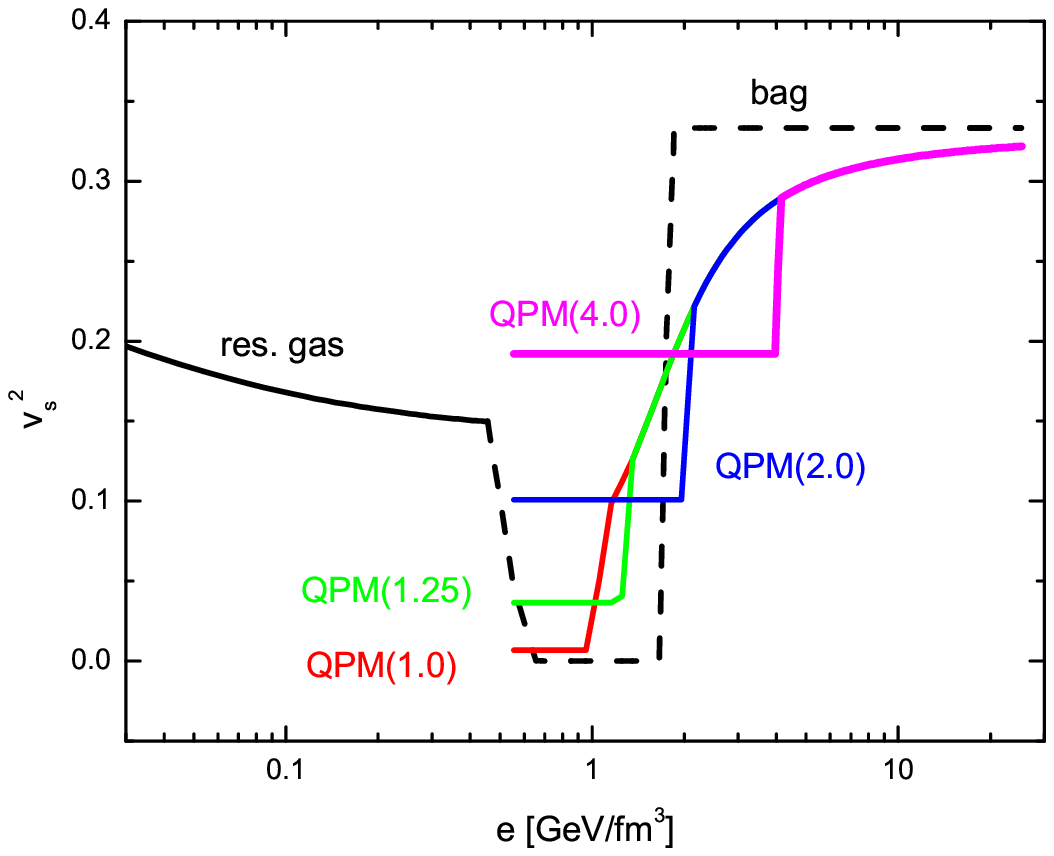}
%  \vspace{4mm}
    \caption{Left panel: Family of EoS's 
    	$p(e)$ labelled in the following as 
	QPM($e_m$) with $e_m=$ 4.0, 2.0, 1.25, 1.0 
	GeV/fm$^3$ (solid curves) combining QPM adjusted to lattice 
	data~\cite{Karsch:2003zq,Karsch:2003vd} 
	and hadron resonance gas at matching point $e_m$. For 
	comparison the bag model EoS (dashed line) is shown. Right panel: corresponding $v_s^2$.}
\end{figure}

\section{Predictions for heavy-ion collisions at LHC}

We concentrate on two extreme EoS's, QPM(4.0) and the bag model EoS being similar 
to QPM(1.0). We calculate transverse momentum spectra and elliptic flow $v_2(p_T)$ using 
the relativistic 
hydrodynamic program package~\cite{Kolb:1999it,Kolb:2000sd} 
with initial conditions for Pb$+$Pb 
collisions at impact parameter $b=5.2$ fm. For the further initial parameters 
required by the program we conservatively guess $s_0=$ 330 fm$^{-3}$, $n_0=$ 0.4 fm$^{-3}$ and 
$\tau_0=$ 0.6 fm/c for initial entropy density, baryon density and time. 
Within the QPM these translate into $e_0=$ 127 GeV/fm$^3$, 
$p_0=$ 42 GeV/fm$^3$ and $T_0=$ 515 MeV. The freeze-out temperature is set $T_{f.o.}=$ 100 MeV. 
In Figure 2, we exhibit our results at midrapidity for various primordial hadron species. 
Striking is the strong radial flow as evident from the flat $p_T$-spectra and a noticeably 
smaller $v_2(p_T)$ than at RHIC in particular at low $p_T$~\cite{Bluhm:2007nu}. 
Details of the Eos in the transition region as mapped out by our family are still visible. 
\begin{figure}[h]
  \includegraphics[scale=0.515]{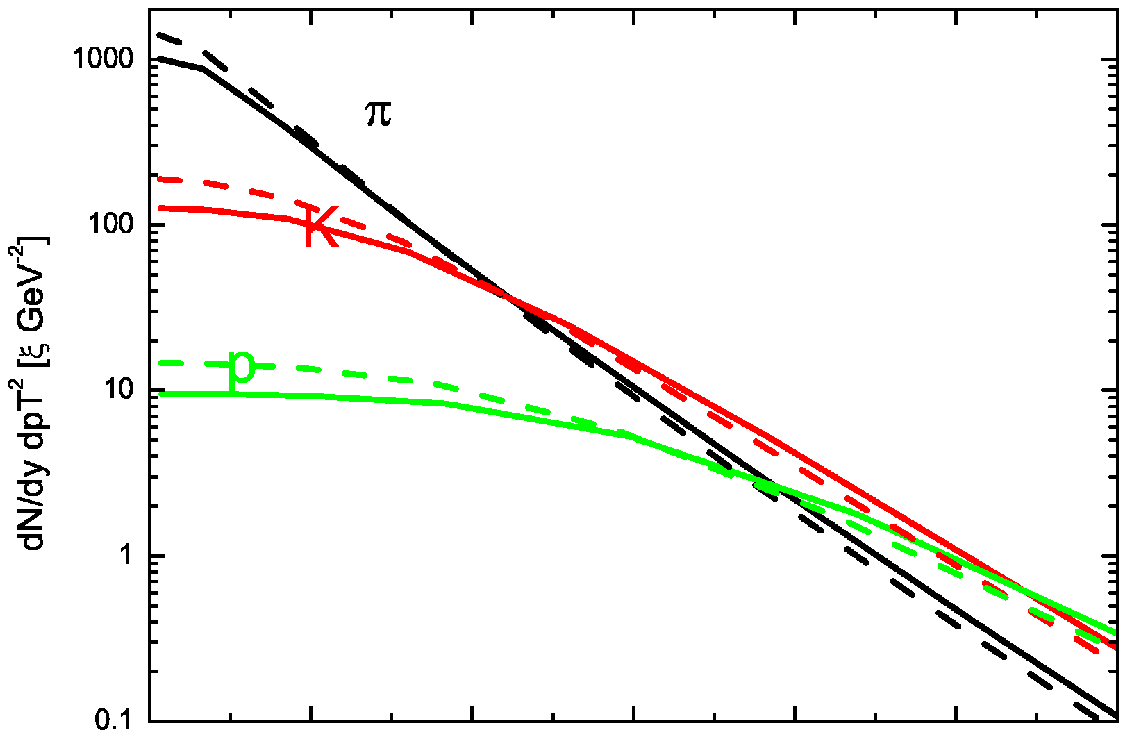}
  \includegraphics[scale=0.515]{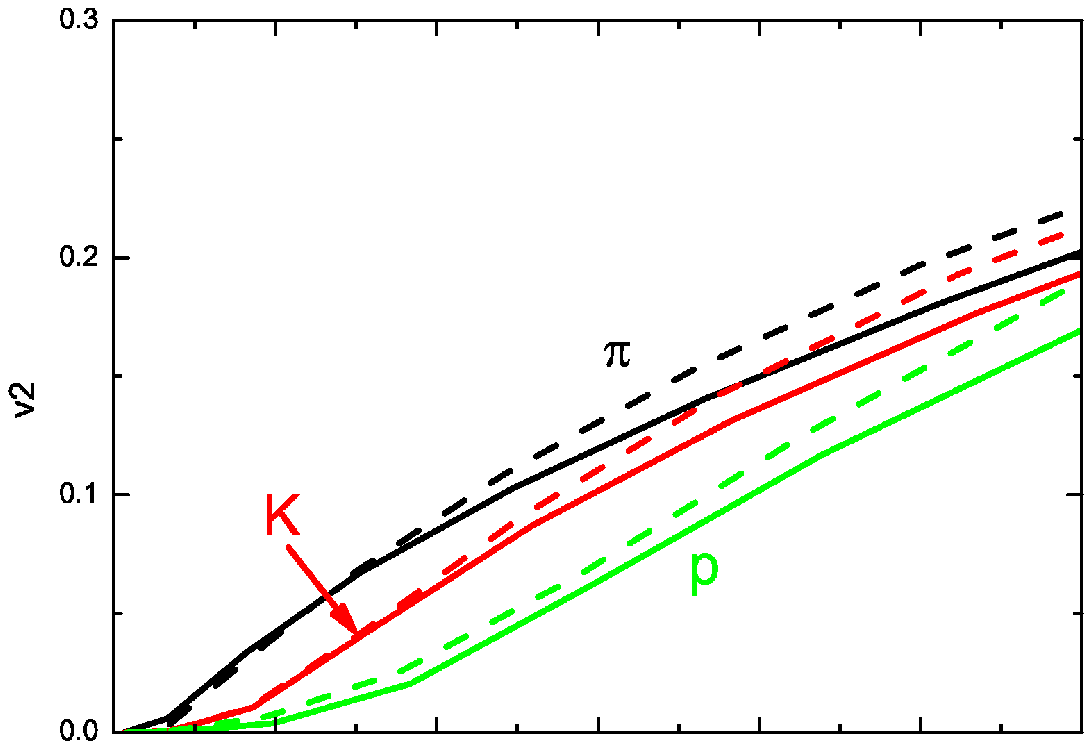}
  \includegraphics[scale=0.518]{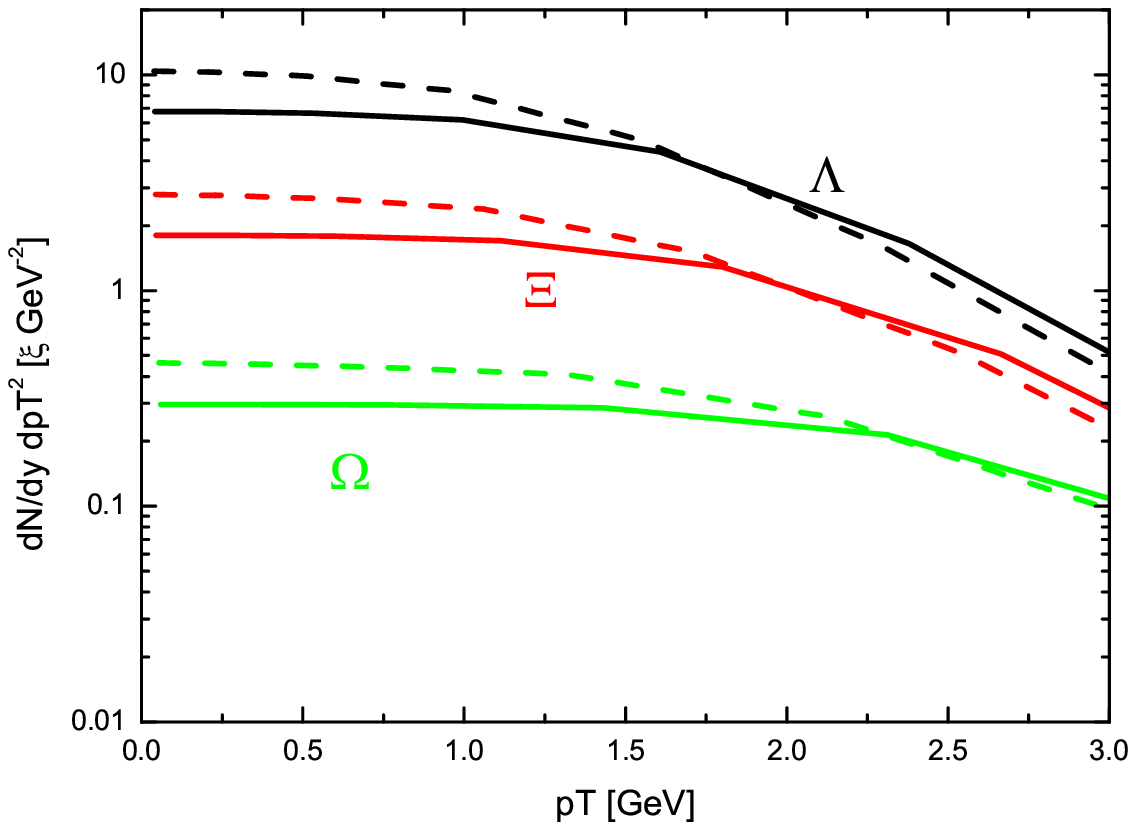}
  \includegraphics[scale=0.515]{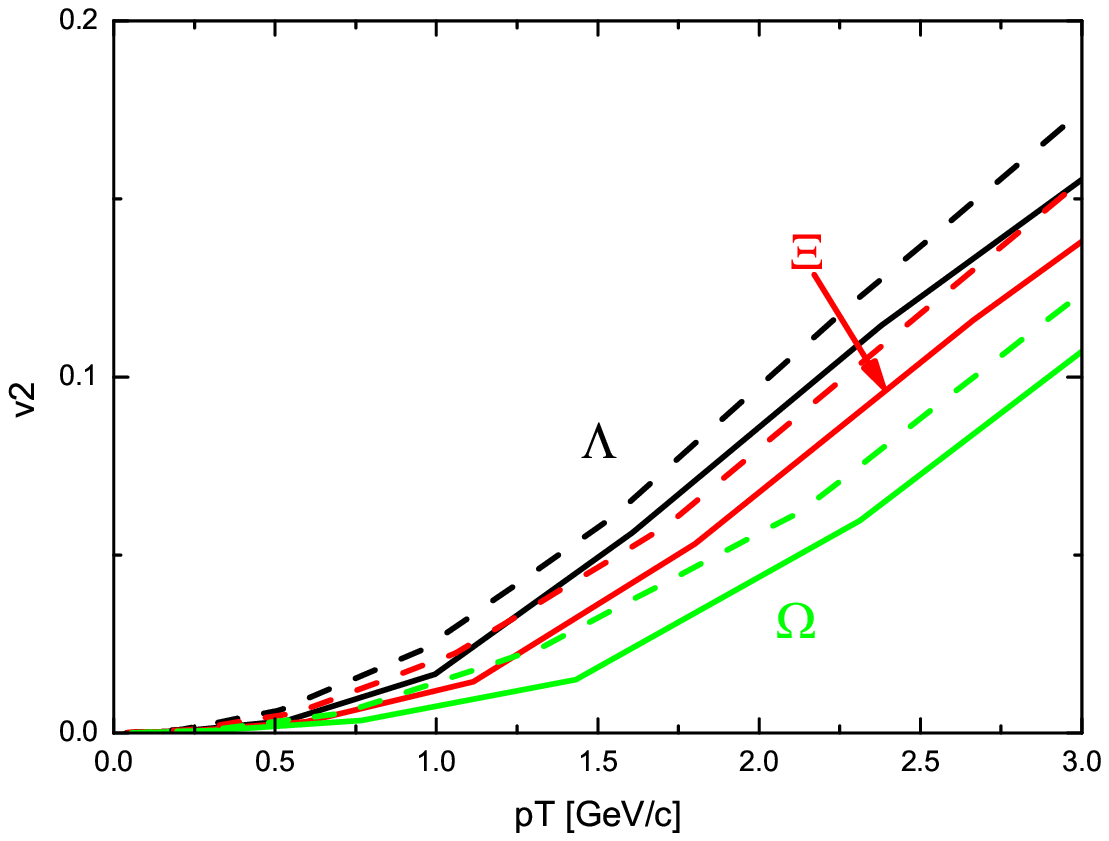}
  \vspace{-7mm}
  \caption{Transverse momentum spectra (left panels) and azimuthal anisotropy (right panels) for 
    directly emitted pions, kaons and protons (upper row) and strange baryons (lower row). Solid 
    and dashed curves are for EoS QPM(4.0) and the bag model EoS, respectively.}
\end{figure}

\newpage

\end{document}